\documentclass[twocolumn,showpacs,preprintnumbers,ams
m a t h , a m s symb]{revtex4} \usepackage{graphicx}

\begin{document}

\preprint{x}

\title{Interaction Effects at Crossings of Spin-Polarised One-Dimensional
Subbands}
\author{A C Graham}
\affiliation{Cavendish Laboratory, Madingley Road,
Cambridge, CB3 OHE, United Kingdom}

\author{K J Thomas}
\affiliation{Cavendish Laboratory, Madingley Road,
Cambridge, CB3 OHE, United Kingdom}

\author{M Pepper}
\affiliation{Cavendish Laboratory, Madingley Road,
Cambridge, CB3 OHE, United Kingdom}

\author{N R Cooper}
\affiliation{Cavendish Laboratory, Madingley Road,
Cambridge, CB3 OHE, United Kingdom}

\author{M Y Simmons$^{\ast}$}
\affiliation{Cavendish Laboratory, Madingley Road,
Cambridge, CB3 OHE, United Kingdom}

\author{D A Ritchie}
\affiliation{Cavendish Laboratory, Madingley Road,
Cambridge, CB3 OHE, United Kingdom}

\title{Interaction Effects at Crossings of Spin-Polarised One-Dimensional
Subbands}

\date{\today}

\begin{abstract}

We report conductance measurements of ballistic
one-dimensional (1D) wires defined in GaAs/AlGaAs heterostructures
in an in-plane magnetic field, $B$.
When the Zeeman energy is equal to the 1D subband energy spacing,
the spin-split subband $N{\uparrow}$ intersects $(N+1){\downarrow}$,
where $N$ is the index of the spin-degenerate 1D subband.
At the crossing of $N=1{\uparrow}$ and $N=2{\downarrow}$
subbands, there is a spontaneous splitting giving rise to an
additional conductance structure evolving from the $1.5(2e^2/h)$ plateau.
With further increase in $B$, the structure develops into a plateau and
lowers to $2e^2/h$. With increasing temperature and magnetic field the
structure shows characteristics of the 0.7 structure.  Our
results suggest that at low densities a spontaneous spin
splitting occurs whenever two 1D subbands of opposite spins
cross.

\end{abstract}

\pacs{71.70.-d, 72.25.Dc, 73.21.Hb, 73.23.Ad}
\maketitle

Studies of ballistic transport in one dimension (1D)
have shown that a spontaneous spin splitting may occur at
zero magnetic field, as indicated
by a conductance structure at $0.7(2e^{2}/h)$, which
drops to $0.5(2e^{2}/h)$ in an in-plane magnetic field\cite{thomas96}.
This so-called 0.7 structure is widely reported in various types of
ballistic 1D wires defined in
GaAs\cite{thomas96,kristensen,reilly,marcus}and Si
heterostructures\cite{bagraev}. In some cases, at very low
electron densities a structure has been observed at
$0.5(2e^{2}/h)$ at zero magnetic field which strengthened
with in-plane magnetic field\cite{thomas00}, indicating a
complete spin polarisation\cite{gold96a,wang96}.

By studying Zeeman splitting of 1D subbands, it was shown that the
0.7 structure is accompanied by an enhancement of the Lande $g$-value
as the 1D subbands are depopulated, and the
energy difference between the spin-split 1D subbands
tends to a finite value at zero magnetic
field\cite{thomas96}.
The conductance plateau at $0.5(2e^2/h)$, either observed in
zero magnetic field or induced by Zeeman effect, rises to
$0.6(2e^{2}/h)$ with increasing temperature\cite{thomas02}.
None of these characteristics can be explained within a
single particle model. As the situation is dynamic, for short
ballistic 1D wires, zero-field spin splitting may not be in
conflict with the theorem of Lieb and Mattis\cite{lieb}.

The discovery of the  0.7 structure in ballistic
1D wires has stimulated
much theoretical work in one-dimension, some of which focused on
zero-field spin polarisation\cite{spivak,bruus,starikov03},
spin density wave formation\cite{reimann}, pairing
of electrons\cite{flambaum}, singlet-triplet
formation\cite{rejec}, Kondo-like
interactions\cite{lindelof,marcus,meir}, and electron-phonon
effects\cite{seelig}. Experimental
studies\cite{thomas96,kristensen,reilly} of the 0.7
structure have, in general, indicated that due to a spin
splitting at zero magnetic field, a complex many-body state
may exist in a ballistic 1D constriction. In order to further
study the role of spin, a strong magnetic field is applied
parallel to a quantum wire to produce large Zeeman splitting
and induce crossings between spin-split 1D subbands\cite{ahrash1}.
In this work, we show that at the crossing of Zeeman-split 1D
subbands of opposite spins and different spatial
wavefunctions, a spontaneous splitting sets in, giving rise
to new conductance structures exhibiting characteristics of
the 0.7 structure; we call this new structure a {\textit{0.7
analogue}}. We have observed such 0.7 analogues in
eight samples, with magnetic field applied in both in-plane
directions.

\begin{figure}
\begin{center}
\includegraphics[scale=0.5]{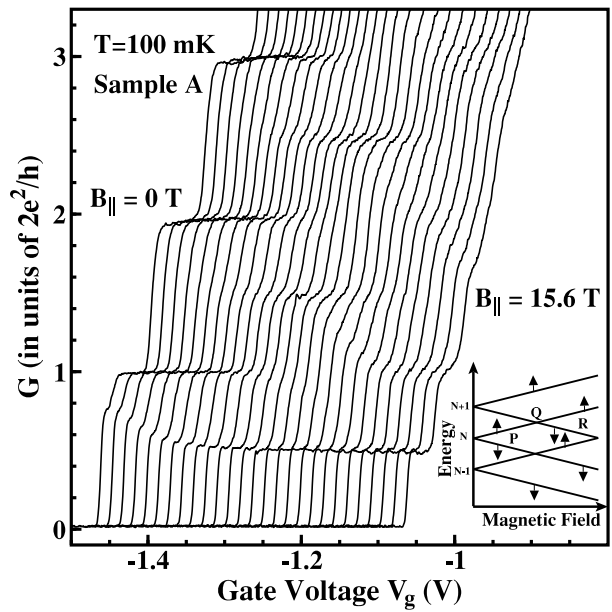}
\end{center}
\caption{Differential conductance, $G(V_g)$, traces at $B_{\parallel}$ incremented in steps of
0.6~T. For clarity, successive traces are offset horizontally. {\textit{Inset}}: Schematic energy
diagram for a linear Zeeman splitting of 1D subbands and subsequent crossings. \label{Fig1}}
\end{figure}

Split-gate devices were defined by electron beam lithography
on a Hall bar etched
from a high mobility GaAs/Al$_x$Ga$_{1-x}$As heterostructure.
Samples A and B used in this work have a length 0.4~$\mu$m and widths
0.6~$\mu$m and 0.5~$\mu$m.
The two-dimensional electron gas (2DEG) formed $292\; $nm below the surface
has a mobility of $1.1\times 10^{6} \;$cm$^{2}$/Vs
and a carrier density of $1.15\times 10^{11}\;$cm$^{-2}$.
Conductance measurements were
performed in a dilution refrigerator using an excitation
voltage of 10 $\mu$V at 77 Hz.  The samples were mounted with
the magnetic field, $B_\parallel$, parallel to the current
direction. By monitoring the Hall voltage, the out-of-plane
misalignment was measured to be less than $0.5\,^{\circ}$.

\begin{figure}
\begin{center}
\includegraphics[width=\columnwidth]{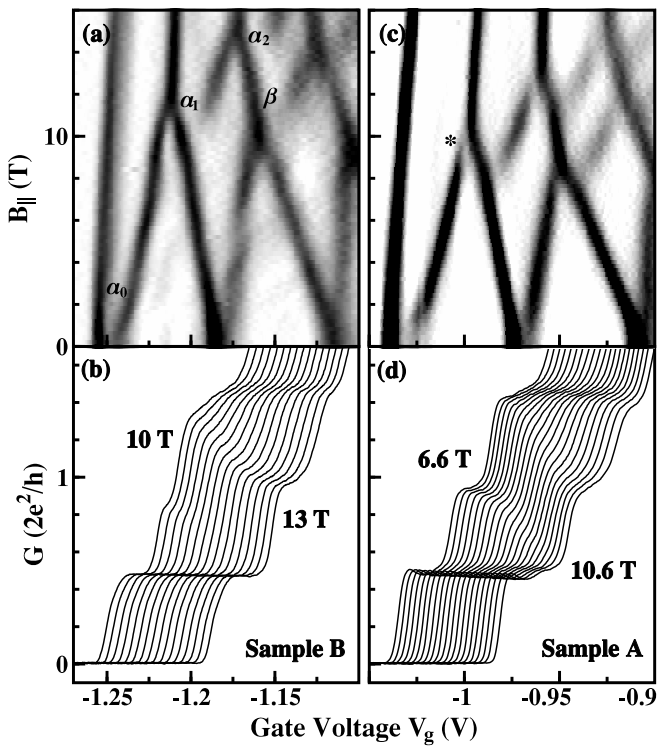}
   \end{center}
\caption{\label{Fig2}(a) Greyscale plot of transconductance,
$dG/dV_{g}$, as a function of $V_{g}$ and
$B_{\parallel}$ for sample B. (b) $G(V_{g})$ traces (offset
horizontally) for $B_{\parallel}=10$~T to $13$~T
incremented by $0.2$~T (sample B, different cool-down).
(c) Greyscale plot as in (a), for sample A.
(d) $G(V_{g})$ traces as in (b), for sample A.}
\end{figure}


Figure \ref{Fig1} shows differential conductance $G=\frac{dI}{dV}$
traces, measured as a function of split-gate voltage $V_{g}$ at
fixed magnetic fields, $B_{\parallel}$. The inset shows a schematic
illustration of linear Zeeman splitting of 1D energy subbands for
a parabolic potential confinement.
The left trace of the main figure shows conductance
plateaus quantised at $N(2e^2/h)$ and the 0.7 structure at
$B_{\parallel}$=0. As $B_{\parallel}$ is incremented to 15.6~T
(right trace), the overall conductance characteristics undergo
three major changes. These correspond to P, Q and R in
Fig.~\ref{Fig1} inset. Firstly,  each spin-degenerate 1D subband
$N$ splits into two, $N{\uparrow}$ and $N{\downarrow}$ (see P in the inset), and new conductance plateaus appear at half-integer values of $2e^2/h$.
Secondly, with further increase of $B_{\parallel}$, the half-integer plateaus
strengthen and integer plateaus weaken. When the Zeeman energy,
$g{\mu}_B B_{\parallel}$, is equal to the subband energy spacing,
$\Delta E_{N,N+1}$, integer plateaus disappear.
This happens when the split levels, for example, $N{\uparrow}$ and
$(N+1){\downarrow}$ converge and pass through a
crossing point (see Q in the inset).
Finally, with further increase of $B_{\parallel}$, half-integer
plateaus weaken and integer plateaus reappear as the
$N{\uparrow}$ and $(N+1){\downarrow}$ diverge again after the
crossing (see R in the inset).
For example, the plateau at $1.5(2e^2/h)$ weakens and the $2e^2/h$ plateau
reappears for $B_{\parallel}>8$~T. However, this is
accompanied by the evolution of
a weak structure from the edge of the $1.5(2e^2/h)$ plateau,
which gradually lowers to $2e^2/h$ and develops into a plateau.
This resembles the evolution of the 0.7 structure to $0.5(2e^2/h)$
with increasing $B_{\parallel}$.
For this reason, we call the structure at the crossing
a 0.7 analogue.
It may be noted that the reappearing $2e^2/h$ plateau now
carries the opposite spin to that before crossing. The lowest subband
1${\downarrow}$, however, does not encounter a crossing;
therefore the plateau at $0.5(2e^2/h)$ is intact, and does not change
its spin.

\begin{figure}
\begin{center}
\includegraphics[width=\columnwidth]{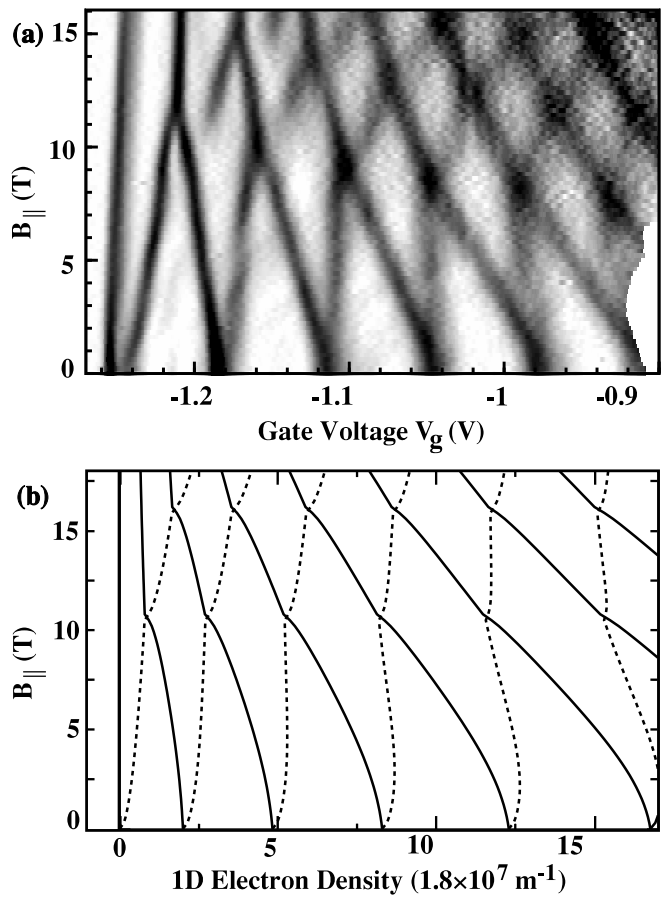}
\end{center}
\caption{\label{Fig3}(a) Greyscale plot of $dG/dV_{g}$ as a
function of $V_{g}$ and $B_{\parallel}$ for sample B, showing the
splitting of five 1D subbands. (b) Calculated Zeeman splitting
with diamagnetic shift of the 1D subbands. The solid
lines represent parallel spin and the dotted lines represent
anti-parallel spin.}
\end{figure}

The evolution of conductance characteristics with the
splitting of 1D subbands in $B_{\parallel}$
can be clearly represented in a grey-scale plot of the transconductance
$dG/dV_g$, obtained by numerical differentiation of $G(V_g)$
characteristics. Figure~\ref{Fig2}(a) shows $dG/dV_g$ plots of sample
B as a function of $B_{\parallel}$ and $V_g$.
White regions represent low transconductance (plateaus in $G(V_g)$)
and the dark thick lines correspond to high transconductance
(transitions between plateaus). Each dark
line splits into two as $B_\parallel$ increases. This can be
interpreted as the splitting of each 1D subband into two subbands of
opposite spins as shown by P in the Fig.~\ref{Fig1} inset.
On the left of Fig.~\ref{Fig2}(a), for $N=1$,
there are two distinct dark lines at $B_{\parallel}=0$.
The white region between these two dark lines represents
the 0.7 structure, marked by ${\alpha_0}$. As the gap between the
$N=1{\downarrow}$ and $N=1{\uparrow}$ widens with
$B_{\parallel}$, the 0.7 structure evolves into a
plateau at $0.5(2e^2/h)$ and the white region (${\alpha_0}$)
in Fig.\ref{Fig2}(a) broadens. At $B_{\parallel}\approx
11$~T, dark lines corresponding to $N=1{\uparrow}$ and
$N=2{\downarrow}$ subbands cross. After the crossing, the
line $N=1{\uparrow}$ shows a discontinuous shift of
${\delta}V_g=23$~mV from the crossing point, marked by
${\alpha_1}$. This discontinuity in the dark line
$N=1{\uparrow}$ corresponds to the appearance of the 0.7
analogue. Figure \ref{Fig2}(b) shows $G(V_g)$ traces 
horizontally) in the vicinity of ${\alpha_1}$ from a
different cool-down of the sample, highlighting the evolution
of the 0.7 analogue from the edge of $1.5(2e^2/h)$ plateau to
$2e^2/h$ with increasing $B_{\parallel}$. It is observed that
the $1.5(2e^2/h)$ plateau, though weakening, remains visible
when the 0.7 analogue evolves.

The discontinuous evolution of the right-moving dark lines
(${\uparrow}$-spin subbands) can also be observed at
the crossing of $N=2{\uparrow}$ with $N=3{\downarrow}$ lines marked
by ${\beta}$, and at the second crossing of $N=1{\uparrow}$ with
$N=3{\downarrow}$ line, marked by ${\alpha_2}$.  Figure \ref{Fig2}(c)
and (d) show results of sample A in a different cool-down from
that of Fig.\ref{Fig1}.
In addition to the splittings at the crossing of peaks
as observed in sample B, in this case there is also a splitting just
before the crossing of $N=1{\uparrow}$ and $N=2{\downarrow}$ peaks,
marked by an asterisk in Fig.\ref{Fig2}(c).
In one of the cool-downs, sample B also showed a weak splitting
before the crossing. It is not clear whether this splitting marked
by asterisk is related to $\alpha_1$, the 0.7 analogue.
It may be observed that this corresponds to a slower rate of
suppression of the $2e^2/h$ plateau (1$\uparrow$-subband) in sample A with
$B_{\parallel}$ as shown in Fig.~\ref{Fig2}(d), compared to
sample B in Fig.~\ref{Fig2}(b).

In the following, we suggest that the main features of our
observations arise from strong electron-electron
interactions. To clarify this, we first describe the expected
behaviour in the case of non-interacting electrons. In
particular, we show that the diamagnetic shifts of
the subband energies in an in-plane magnetic field have only
a very limited influence on the most important features of
our observations.

Figure~\ref{Fig3}(a) shows a greyscale plot of sample B with
many occupied 1D subbands, part of which is shown in
Fig.~\ref{Fig2}(a). Figure.~\ref{Fig3}(b) shows the positions
of the calculated transconductance peaks as a function of
electron density and $B_{\parallel}$ for a model of
non-interacting electrons in an infinite 1D wire.  This model
includes the diamagnetic effects with $B_{\parallel}$,
assuming parabolic confinement in the transverse and vertical
(quantum well) directions with subband spacings of 1.85~meV
and 15~meV \cite{ando} respectively. A $g$-value of 1.9 is
used in this model in accordance with the value measured at
low $B_{\parallel}$ in our samples.

If one assumes that the electron density in the wire is
linearly related to $V_{g}$, then Figs.~\ref{Fig3}(a)
and (b) can be compared directly.  Clearly
the model of non-interacting electrons accounts well for
the general trends in the evolution of the transconductance
peaks with $V_g$ and $B_{\parallel}$.
However, the model cannot capture the appearance of
discontinuities in the positions of the transconducance peaks
at the crossings, $\alpha_1$, $\alpha_2$, and $\beta$ in
Fig.~\ref{Fig2}. As we have described above, these are the
regions where the conductance displays the 0.7 analogues.

\begin{figure}
\begin{center}
\includegraphics[scale=0.5]{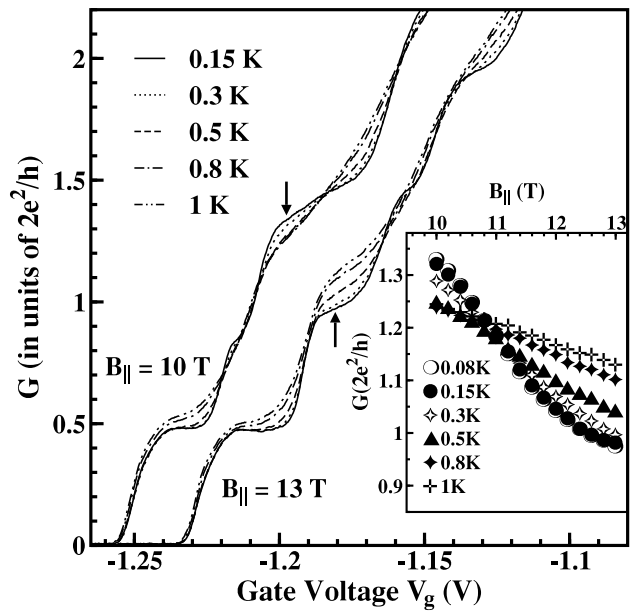}
\end{center}
\caption{\label{Fig4} Temperature dependence of $G(V_{g})$
for sample B (same cool-down as Fig.~\ref{Fig2}(b)) at 10 T
and 13 T. {\textit{Inset}}: The height of the 0.7 analogue as
a function of $B_{\parallel}$ for various temperatures.}
\end{figure}

Figure \ref{Fig4} shows the temperature dependence of the 0.7
analogue at the first crossing of 1$\uparrow$ and
2$\downarrow$ subbands. A defining characteristic of the
0.7 structure is its unusual temperature dependence. In
addition, at low temperatures, the 0.7 structure becomes
well-defined only at low electron densities\cite{thomas00}.
For a higher 1D density, a higher temperature (typically,
$T\approx 1$~K in GaAs split-gate devices\cite{thomas96}) is
required to induce the 0.7 structure\cite{kristensen}. At
high $B_{\parallel}$, it is well-known that the 0.7 structure
develops into a plateau at $0.5(2e^2/h)$; however with an
increase in $T$, the plateau rises to
$0.6(2e^2/h)$\cite{thomas02}. These trends in the temperature
dependence of the 0.7 structure and $0.5(2e^2/h)$ plateau are
also observed in the new 0.7 analogue structure at
$B_{\parallel}=10$~T and $2e^2/h$ plateau at
$B_{\parallel}=13$~T. Figure \ref{Fig4} shows that, at
$B_{\parallel}=10$~T, the 0.7 analogue present just below
$1.5(2e^{2}/h)$ (shown by a down-arrow) drops as $T$ rises;
but at $B_{\parallel}=13$~T, the plateau at $2e^{2}/h$ (shown
by an up-arrow) rises with increasing $T$. In the inset of
Fig.~\ref{Fig4}, conductance of the 0.7 analogue is plotted
as a function of $B_{\parallel}$ for a range of temperatures.
This data compares well to the temperature dependence of the
0.7 structure \cite{thomas02}, showing the crossover of $G$
as a function of $T$ at a characteristic $B_{\parallel}$.


  Non-quantised conductance structures can be due to a
change in the transmission probability caused by scattering
or a many-body effect in the 1D channel. We have observed 0.7
analogues in eight samples, and they are independent of
cool-downs, and occur only at the crossing of spin-split
subbands of opposite spins. Due to the high reproducibility
of the 0.7 analogues, a disorder-induced scattering effect
can be discounted.

When two energy levels are brought together, an anticrossing
may occur. However, this depends on the symmetry of the two
wavefunctions. In our case, the two 1D levels that cross in
$B_{\parallel}$ have different spins and subband
indices; therefore such anticrossings should be very weak.
Experimentally, we do not observe anticrossings of 1D
subbands, rather a gap forms abruptly after the crossing.
As in the case of the 0.7 structure, we believe that the
new 0.7 analogue is a consequence of strong exchange
interactions. In the former case, there is a lifting of
the zero-field spin-degeneracy, whereas in the latter case, the
degeneracy at the crossing point is lifted.
We can quantify the strength
of the exchange interactions by measuring the gate voltage
splitting at $\alpha_1$.
From dc source-drain bias calibration\cite{patel91a},
$\alpha_1$ is measured to be 0.5~meV, which is a third of the
subband spacing ($\Delta{E_{1,2}}=1.6$~meV) at zero magnetic
field. 

Given the strong similarity between the 0.7 analogue
and the 0.7 structure, we consider whether theories for
the 0.7 structure could apply to our results.
A recently proposed electron-phonon scattering mechanism for
the 0.7 structure\cite{seelig} cannot account for the observed
0.7 analogue. Consider Fig. \ref{Fig2}(b): if the
strong 0.7 analogue in these traces were the result of a
conductance suppression caused by electron-phonon scattering,
then one should expect at least as strong a suppression below
the spin-polarised $0.5(2e^{2}/h)$ plateau; there is
no such indication.

The behaviour of the 0.7 analogue does not seem to be consistent with
the most simple extension of a proposed ``Kondo model" for the 0.7
structure\cite{meir} to the present situation -- in which a
Kondo impurity forms from quasibound states of the 1$\uparrow$
and 2$\downarrow$ levels which become degenerate
at some non-zero $B_\parallel$.
In this model, one would expect the 0.7 analogue feature to
fall onto the $2e^2/h$
plateau as $B_\parallel$ is increased or decreased away from
the point of degeneracy in {\it either} direction. In contrast,
the 0.7 analogue evolves asymmetrically about the midpoint of the
crossing. A detailed analysis within the Kondo model will be
considered in a later publication.

To conclude, we have observed the crossings of spin-split
1D subbands of different spins and spatial wavefunctions in a
1D electron gas. At crossings, there is a
spontaneous splitting giving rise to new conductance
structures. There are no indications of anticrossings, but an
energy splitting may occur whenever two 1D subbands of
opposite spin are nearly degenerate. The magnetic field and
temperature dependences show that the new structures strongly
resemble the zero-field 0.7 structure. We believe that these
0.7 analogue structures may provide the key to a fuller
understanding of the role of electron-electron interactions
in ballistic 1D wires.

We thank J. T. Nicholls, D. Khmelnitskii, V. Tripathi, and C. J. B. Ford
for useful discussions. This work was supported by  EPSRC, UK.
KJT acknowledges support from the Royal Society.

$^{\ast}$Current address: University of New South Wales, School of Physics, Sydney, NSW 2052, Australia.


\begin{thebibliography}{25}
\expandafter\ifx\csname natexlab\endcsname\relax\def\natexlab#1{#1}\fi
\expandafter\ifx\csname bibnamefont\endcsname\relax
  \def\bibnamefont#1{#1}\fi
\expandafter\ifx\csname bibfnamefont\endcsname\relax
  \def\bibfnamefont#1{#1}\fi
\expandafter\ifx\csname citenamefont\endcsname\relax
  \def\citenamefont#1{#1}\fi
\expandafter\ifx\csname url\endcsname\relax
  \def\url#1{\texttt{#1}}\fi
\expandafter\ifx\csname urlprefix\endcsname\relax\def\urlprefix{URL }\fi
\providecommand{\bibinfo}[2]{#2}
\providecommand{\eprint}[2][]{\url{#2}}

\bibitem[{\citenamefont{Thomas et~al.}(1996)\citenamefont{Thomas, Nicholls,
  Simmons, Pepper, Mace, and Ritchie}}]{thomas96}
\bibinfo{author}{\bibfnamefont{K.~J.} \bibnamefont{Thomas}},
  \bibinfo{author}{\bibfnamefont{J.~T.} \bibnamefont{Nicholls}},
  \bibinfo{author}{\bibfnamefont{M.~Y.} \bibnamefont{Simmons}},
  \bibinfo{author}{\bibfnamefont{M.}~\bibnamefont{Pepper}},
  \bibinfo{author}{\bibfnamefont{D.~R.} \bibnamefont{Mace}}, \bibnamefont{and}
  \bibinfo{author}{\bibfnamefont{D.~A.} \bibnamefont{Ritchie}},
  \bibinfo{journal}{Phys.\ Rev.\ Lett.} \textbf{\bibinfo{volume}{77}},
  \bibinfo{pages}{135} (\bibinfo{year}{1996}).


\bibitem[{\citenamefont{Kristensen et~al.}(1998)\citenamefont{Kristensen,
  Lindelof, Jensen, Zaffalon, Hollingbery, Pedersen, Nygard, Bruus, Reimann,
  S{\"o}renson et~al.}}]{kristensen}
\bibinfo{author}{\bibfnamefont{A.}~\bibnamefont{Kristensen}},
  \bibinfo{author}{\bibfnamefont{P.~E.} \bibnamefont{Lindelof}},
  \bibinfo{author}{\bibfnamefont{J.~B.} \bibnamefont{Jensen}},
  \bibinfo{author}{\bibfnamefont{M.}~\bibnamefont{Zaffalon}},
  \bibinfo{author}{\bibfnamefont{J.}~\bibnamefont{Hollingbery}},
  \bibinfo{author}{\bibfnamefont{S.~W.} \bibnamefont{Pedersen}},
  \bibinfo{author}{\bibfnamefont{J.}~\bibnamefont{Nygard}},
  \bibinfo{author}{\bibfnamefont{H.}~\bibnamefont{Bruus}},
  \bibinfo{author}{\bibfnamefont{S.~M.} \bibnamefont{Reimann}},
  \bibinfo{author}{\bibfnamefont{C.~B.} \bibnamefont{S{\"o}renson}},
  \bibnamefont{et~al.}, \bibinfo{journal}{Physica B}
  \textbf{\bibinfo{volume}{251}}, \bibinfo{pages}{180} (\bibinfo{year}{1998}).

\bibitem[{\citenamefont{Reilly et~al.}(2001)\citenamefont{Reilly, Facer,
  Dzurak, Kane, Clark, Stiles, Clark, Hamilton, O'Brien, Lumpkin
  et~al.}}]{reilly}
\bibinfo{author}{\bibfnamefont{D.~J.} \bibnamefont{Reilly}},
  \bibinfo{author}{\bibfnamefont{G.~R.} \bibnamefont{Facer}},
  \bibinfo{author}{\bibfnamefont{A.~S.} \bibnamefont{Dzurak}},
  \bibinfo{author}{\bibfnamefont{B.~E.} \bibnamefont{Kane}},
  \bibinfo{author}{\bibfnamefont{R.~G.} \bibnamefont{Clark}},
  \bibinfo{author}{\bibfnamefont{P.~J.} \bibnamefont{Stiles}},
  \bibinfo{author}{\bibfnamefont{R.~G.} \bibnamefont{Clark}},
  \bibinfo{author}{\bibfnamefont{A.~R.} \bibnamefont{Hamilton}},
  \bibinfo{author}{\bibfnamefont{J.~L.} \bibnamefont{O'Brien}},
  \bibinfo{author}{\bibfnamefont{N.~E.} \bibnamefont{Lumpkin}},
  \bibnamefont{et~al.}, \bibinfo{journal}{Phys.\ Rev.\ B}
  \textbf{\bibinfo{volume}{63}}, \bibinfo{pages}{121311}
  (\bibinfo{year}{2001}).

\bibitem[{\citenamefont{Cronenwett et~al.}(2002)\citenamefont{Cronenwett,
  Lynch, Goldhaber-Gordon, Kouwenhoven, Marcus, Hirose, Wingreen, and
  Umansky}}]{marcus}
\bibinfo{author}{\bibfnamefont{S.~M.} \bibnamefont{Cronenwett}},
  \bibinfo{author}{\bibfnamefont{H.~J.} \bibnamefont{Lynch}},
  \bibinfo{author}{\bibfnamefont{D.}~\bibnamefont{Goldhaber-Gordon}},
  \bibinfo{author}{\bibfnamefont{L.~P.} \bibnamefont{Kouwenhoven}},
  \bibinfo{author}{\bibfnamefont{C.~M.} \bibnamefont{Marcus}},
  \bibinfo{author}{\bibfnamefont{K.}~\bibnamefont{Hirose}},
  \bibinfo{author}{\bibfnamefont{N.~S.} \bibnamefont{Wingreen}},
  \bibnamefont{and} \bibinfo{author}{\bibfnamefont{V.}~\bibnamefont{Umansky}},
  \bibinfo{journal}{Phys.\ Rev.\ Lett.} \textbf{\bibinfo{volume}{88}},
  \bibinfo{pages}{226805} (\bibinfo{year}{2002}).

\bibitem[{\citenamefont{Bagraev et~al.}(2002)\citenamefont{Bagraev, Buravlev,
  Klyachkin, Malyarenko, Gehlhoff, Ivanov, and Shelykh}}]{bagraev}
\bibinfo{author}{\bibfnamefont{N.~T.} \bibnamefont{Bagraev}},
  \bibinfo{author}{\bibfnamefont{A.~D.} \bibnamefont{Buravlev}},
  \bibinfo{author}{\bibfnamefont{L.~E.} \bibnamefont{Klyachkin}},
  \bibinfo{author}{\bibfnamefont{A.~M.} \bibnamefont{Malyarenko}},
  \bibinfo{author}{\bibfnamefont{W.}~\bibnamefont{Gehlhoff}},
  \bibinfo{author}{\bibfnamefont{V.~K.} \bibnamefont{Ivanov}},
  \bibnamefont{and} \bibinfo{author}{\bibfnamefont{I.~A.}
  \bibnamefont{Shelykh}}, \bibinfo{journal}{Semiconductors}
  \textbf{\bibinfo{volume}{36}}, \bibinfo{pages}{439} (\bibinfo{year}{2002}).

\bibitem[{\citenamefont{Thomas et~al.}(2000)\citenamefont{Thomas, Nicholls,
  Pepper, Tribe, Simmons, and Ritchie}}]{thomas00}
\bibinfo{author}{\bibfnamefont{K.~J.} \bibnamefont{Thomas}},
  \bibinfo{author}{\bibfnamefont{J.~T.} \bibnamefont{Nicholls}},
  \bibinfo{author}{\bibfnamefont{M.}~\bibnamefont{Pepper}},
  \bibinfo{author}{\bibfnamefont{W.~R.} \bibnamefont{Tribe}},
  \bibinfo{author}{\bibfnamefont{M.~Y.} \bibnamefont{Simmons}},
  \bibnamefont{and} \bibinfo{author}{\bibfnamefont{D.~A.}
  \bibnamefont{Ritchie}}, \bibinfo{journal}{Phys.\ Rev.\ B}
  \textbf{\bibinfo{volume}{61}}, \bibinfo{pages}{13365} (\bibinfo{year}{2000}).

\bibitem[{\citenamefont{Wang and Berggren}(1996)}]{wang96}
\bibinfo{author}{\bibfnamefont{C.-K.} \bibnamefont{Wang}} \bibnamefont{and}
  \bibinfo{author}{\bibfnamefont{K.-F.} \bibnamefont{Berggren}},
  \bibinfo{journal}{Phys.\ Rev.\ B} \textbf{\bibinfo{volume}{54}},
  \bibinfo{pages}{14257} (\bibinfo{year}{1996}).

\bibitem[{\citenamefont{Gold and Calmels}(1996)}]{gold96a}
\bibinfo{author}{\bibfnamefont{A.}~\bibnamefont{Gold}} \bibnamefont{and}
  \bibinfo{author}{\bibfnamefont{L.}~\bibnamefont{Calmels}},
  \bibinfo{journal}{Phil.\ Mag.\ Lett.} \textbf{\bibinfo{volume}{74}},
  \bibinfo{pages}{33} (\bibinfo{year}{1996}).

\bibitem[{\citenamefont{Thomas et~al.}(2002)\citenamefont{Thomas, Nicholls,
  Pepper, Simmons, Mace, and Ritchie}}]{thomas02}
\bibinfo{author}{\bibfnamefont{K.~J.} \bibnamefont{Thomas}},
  \bibinfo{author}{\bibfnamefont{J.~T.} \bibnamefont{Nicholls}},
  \bibinfo{author}{\bibfnamefont{M.}~\bibnamefont{Pepper}},
  \bibinfo{author}{\bibfnamefont{M.~Y.} \bibnamefont{Simmons}},
  \bibinfo{author}{\bibfnamefont{D.~R.} \bibnamefont{Mace}}, \bibnamefont{and}
  \bibinfo{author}{\bibfnamefont{D.~A.} \bibnamefont{Ritchie}},
  \bibinfo{journal}{Physica E} \textbf{\bibinfo{volume}{12}},
  \bibinfo{pages}{708} (\bibinfo{year}{2002}).

\bibitem[{\citenamefont{Lieb and Mattis}(1962)}]{lieb}
\bibinfo{author}{\bibfnamefont{E.}~\bibnamefont{Lieb}} \bibnamefont{and}
  \bibinfo{author}{\bibfnamefont{D.}~\bibnamefont{Mattis}},
  \bibinfo{journal}{Phys.\ Rev.} \textbf{\bibinfo{volume}{125}},
  \bibinfo{pages}{164} (\bibinfo{year}{1962}).

\bibitem[{\citenamefont{Spivak and Zhou}(2000)}]{spivak}
\bibinfo{author}{\bibfnamefont{B.}~\bibnamefont{Spivak}} \bibnamefont{and}
  \bibinfo{author}{\bibfnamefont{F.}~\bibnamefont{Zhou}},
  \bibinfo{journal}{Phys.\ Rev.\ B} \textbf{\bibinfo{volume}{61}},
  \bibinfo{pages}{16730} (\bibinfo{year}{2000}).

\bibitem[{\citenamefont{Bruus et~al.}(2001)\citenamefont{Bruus, Cheianov, and
  Flensberg}}]{bruus}
\bibinfo{author}{\bibfnamefont{H.}~\bibnamefont{Bruus}},
  \bibinfo{author}{\bibfnamefont{V.~V.} \bibnamefont{Cheianov}},
  \bibnamefont{and}
  \bibinfo{author}{\bibfnamefont{K.}~\bibnamefont{Flensberg}},
  \bibinfo{journal}{Physica E} \textbf{\bibinfo{volume}{10}},
  \bibinfo{pages}{97} (\bibinfo{year}{2001}).

\bibitem[{\citenamefont{Starikov et~al.}(2003)}]{starikov03}
\bibinfo{author}{\bibfnamefont{A.~A.}
\bibnamefont{Starikov}},
\bibinfo{author}{\bibfnamefont{I.~I.}
\bibnamefont{Yakimenko}}
\bibnamefont{and}
\bibinfo{author}{\bibfnamefont{K.-F.}
\bibnamefont{Berggren}},
\bibinfo{journal}{Phys.\ Rev.\ B}
\textbf{\bibinfo{volume}{67}},   \bibinfo{pages}{235319}
(\bibinfo{year}{2003}).




\bibitem[{\citenamefont{Reimann et~al.}(1999)\citenamefont{Reimann, Koskinen,
  and Manninen}}]{reimann}
\bibinfo{author}{\bibfnamefont{S.~M.} \bibnamefont{Reimann}},
  \bibinfo{author}{\bibfnamefont{M.}~\bibnamefont{Koskinen}}, \bibnamefont{and}
  \bibinfo{author}{\bibfnamefont{M.}~\bibnamefont{Manninen}},
  \bibinfo{journal}{Phys.\ Rev.\ B} \textbf{\bibinfo{volume}{59}},
  \bibinfo{pages}{1613} (\bibinfo{year}{1999}).

\bibitem[{\citenamefont{Flambaum and Kuchiev}(2000)}]{flambaum}
\bibinfo{author}{\bibfnamefont{V.~V.} \bibnamefont{Flambaum}} \bibnamefont{and}
  \bibinfo{author}{\bibfnamefont{M.~Y.} \bibnamefont{Kuchiev}},
  \bibinfo{journal}{Phys.\ Rev.\ B} \textbf{\bibinfo{volume}{61}},
  \bibinfo{pages}{7869} (\bibinfo{year}{2000}).

\bibitem[{\citenamefont{Rejec et~al.}(2000)\citenamefont{Rejec, Ramsak, and
  Jefferson}}]{rejec}
\bibinfo{author}{\bibfnamefont{T.}~\bibnamefont{Rejec}},
  \bibinfo{author}{\bibfnamefont{A.}~\bibnamefont{Ramsak}}, \bibnamefont{and}
  \bibinfo{author}{\bibfnamefont{J.~H.} \bibnamefont{Jefferson}},
  \bibinfo{journal}{Phys.\ Rev.\ B} \textbf{\bibinfo{volume}{62}},
  \bibinfo{pages}{12985} (\bibinfo{year}{2000}).

\bibitem[{\citenamefont{Meir et~al.}(2002)\citenamefont{Meir, Hirose, and
  Wingreen}}]{meir}
\bibinfo{author}{\bibfnamefont{Y.}~\bibnamefont{Meir}},
  \bibinfo{author}{\bibfnamefont{K.}~\bibnamefont{Hirose}}, \bibnamefont{and}
  \bibinfo{author}{\bibfnamefont{N.~S.} \bibnamefont{Wingreen}},
  \bibinfo{journal}{Phys.\ Rev.\ Lett.} \textbf{\bibinfo{volume}{89}},
  \bibinfo{pages}{196802} (\bibinfo{year}{2002}).

\bibitem[{\citenamefont{Lindelof}(2001)}]{lindelof}
\bibinfo{author}{\bibfnamefont{P.~E.} \bibnamefont{Lindelof}},
  \bibinfo{journal}{Proc. SPIE} \textbf{\bibinfo{volume}{4415}},
  \bibinfo{pages}{12} (\bibinfo{year}{2001}).

\bibitem[{\citenamefont{Seelig and Matveev}(2002)}]{seelig}
\bibinfo{author}{\bibfnamefont{G.}~\bibnamefont{Seelig}}
\bibnamefont{and}
\bibinfo{author}{\bibfnamefont{K.~A.} \bibnamefont{Matveev}},
\bibinfo{journal}{Phys.\ Rev.\ Lett.}
\textbf{\bibinfo{volume}{90}},
\bibinfo{pages}{176804}(\bibinfo{year}{2003}).


\bibitem[{\citenamefont{Daneshvar et~al.}(1997)\citenamefont{Daneshvar, Ford,
  Hamilton, Simmons, Pepper, and Ritchie}}]{ahrash1}
\bibinfo{author}{\bibfnamefont{A.~J.} \bibnamefont{Daneshvar}},
  \bibinfo{author}{\bibfnamefont{C.~J.~B.} \bibnamefont{Ford}},
  \bibinfo{author}{\bibfnamefont{A.~R.} \bibnamefont{Hamilton}},
  \bibinfo{author}{\bibfnamefont{M.~Y.} \bibnamefont{Simmons}},
  \bibinfo{author}{\bibfnamefont{M.}~\bibnamefont{Pepper}}, \bibnamefont{and}
  \bibinfo{author}{\bibfnamefont{D.~A.} \bibnamefont{Ritchie}},
  \bibinfo{journal}{Phys.\ Rev.\ B} \textbf{\bibinfo{volume}{55}},
  \bibinfo{pages}{13409} (\bibinfo{year}{1997}).

\bibitem[{\citenamefont{Ando}(1982)}]{ando}
\bibinfo{author}{\bibfnamefont{T.}~\bibnamefont{Ando}}, \bibinfo{journal}{J.\
  Phys.\ Soc.\ Jpn.} \textbf{\bibinfo{volume}{51}}, \bibinfo{pages}{3893}
  (\bibinfo{year}{1982}).

\bibitem[{\citenamefont{Patel et~al.}(1991)\citenamefont{Patel, Nicholls,
  Mart\'{\i}n-Moreno, Pepper, Frost, Ritchie, and Jones}}]{patel91a}
\bibinfo{author}{\bibfnamefont{N.~K.} \bibnamefont{Patel}},
  \bibinfo{author}{\bibfnamefont{J.~T.} \bibnamefont{Nicholls}},
  \bibinfo{author}{\bibfnamefont{L.}~\bibnamefont{Mart\'{\i}n-Moreno}},
  \bibinfo{author}{\bibfnamefont{M.}~\bibnamefont{Pepper}},
  \bibinfo{author}{\bibfnamefont{J.~E.~F.} \bibnamefont{Frost}},
  \bibinfo{author}{\bibfnamefont{D.~A.} \bibnamefont{Ritchie}},
  \bibnamefont{and} \bibinfo{author}{\bibfnamefont{G.~A.~C.}
  \bibnamefont{Jones}}, \bibinfo{journal}{Phys.\ Rev.\ B}
  \textbf{\bibinfo{volume}{44}}, \bibinfo{pages}{13549} (\bibinfo{year}{1991}).

\end{thebibliography}
\end{document}